\documentclass[aps,twocolumn,floatfix,showpacs,amssymb]{revtex4}
\usepackage{graphicx,bm}
\usepackage{hyperref}
\usepackage{amsmath}
\usepackage{mathbbol}
\usepackage{color}

\begin{document}
\title{Ultracold-atom collisions in atomic waveguides: A two-channel analysis}
\author{Tom Kristensen$^{1,2}$ and Ludovic Pricoupenko$^{1,2}$}
\affiliation
{
1- Sorbonne Universit\'{e}s, UPMC Univ Paris 06, UMR 7600, Laboratoire de Physique Th\'{e}orique de la Mati\`{e}re Condens\'{e}e, F-75005, Paris, France\\
2- CNRS, UMR 7600, Laboratoire de Physique Th\'{e}orique de la Mati\`{e}re Condens\'{e}e, F-75005, Paris, France
}
\date{\today}


\begin{abstract}
Low dimensional behavior of two ultra-cold atoms trapped in two- and one-dimensional waveguides is investigated in the vicinity of a magnetic Feshbach resonance. A quantitative two-channel model for the Feshbach mechanism is used, allowing an exhaustive analysis of low-dimensional resonant scattering behavior and of the confinement induced bound states. The role of the different parameters of the resonance is depicted in this context. Results are compared with the ones of the zero-range approach. The relevance of the effective range approximation in low dimensions is studied. Examples of known resonances are used to illustrate the bound state properties.  
\end{abstract}

\maketitle

\section{INTRODUCTION}

The use of magnetic Feshbach resonances and highly anisotropic traps offers the possibility to achieve strongly correlated dilute ultra-cold atomic gases. Feshbach resonances permit the tuning of the scattering length (denoted ${a}$) for two colliding atoms by using an external magnetic field \cite{Chi10}. The standard expression for the scattering length as a function of the external magnetic field ${B}$ for a Feshbach resonance (FR) located at the magnetic field ${B=B_0}$ is
\begin{equation}
a=a_{\rm bg} \left(1 - \frac{\Delta  B}{B-B_0} \right) ,
\label{eq:a}
\end{equation}
where ${a_{\rm bg}}$ is the background scattering length, i.e., the scattering length for large magnetic detuning, and ${\Delta {B}}$ is the magnetic width of the FR. Feshbach resonances can be used for many atomic species and depending on the value of the magnetic field, the ultracold gas in an optical trap can be studied in all the scattering regimes from very small interactions (for a small value of the scattering length) to the resonant regime (for a large value of the scattering length). The fine control of the trapping frequencies permits also to reach geometries where the one-dimensional (1D) or the two-dimensional (2D) behavior of the ultracold gas can be observed \cite{QGLD,Blo08}. In the limit of a purely 1D (respectively 2D) geometry, atoms are free to move along one (respectively two) direction(s) whereas the system is frozen in the transverse direction(s). These configurations can be reached by using a 2D (or 1D) harmonic trap which plays the role of an atomic waveguide. This transverse trap leads to confinement induced resonances in low dimensional scattering, predicted in Refs.~\cite{Ols98,Pet00} and observed in Ref.~\cite{Hal10}.

For a gas of density ${n}$ in a trap of atomic frequency ${\omega_\perp}$ and transverse length ${a_\perp}$, the $D$-dimensional behavior at temperature $T$ is reached in the limit where ${k_B T \ll \hbar \omega_\perp}$ and ${n a_\perp^3 \ll 1 }$. The degenerate regime is achieved for atomic densities $n$ larger than or of the order of ${a_\perp^{D-3}/\lambda_T^D}$ where ${\lambda_T}$ is the de Broglie wavelength. The limit of Tonks-Girardeau and super Tonks-Girardeau \cite{Gir60,Ast05} are two celebrated examples of highly correlated 1D phase which have been achieved thanks to ultra-cold atoms experiments \cite{Par04,Kin04a,Hal09}. In the 2D geometry the interplay between the Bose-Einstein condensate (BEC) and the Berezinskii-Kosterlitz-Thouless (BKT) transitions has been subject to intensive experimental studies \cite{Had06,Had13}. 

Aside from the properties of the systems in the strict 1D or 2D limits, i.e., in the monomode regime of the atomic waveguide, depending on the temperature, the atomic density, and the trap parameters, the study of the transition from three-dimensional (3D) to the low-dimensional physics has also its own fundamental interest. In ongoing studies, the quasi-1D or quasi-2D nature of the system, i.e., the population of the transverse modes of the atomic waveguide as a function of the relative energy in two-body processes is a relevant issue in the few- and many-body problems. For example, this permits us to understand the emergence of Efimov physics in confined geometries \cite{Mor05a,Lev14} or also the contribution of transverse modes in the experimental  studies of the BKT transition \cite{Hol08,Pli11}. 

Due to the crucial relevance of binary collisions in atomic waveguides, this paper is aimed at presenting a detailed analysis of the two-body problem for atoms in 1D and 2D atomic waveguides in the vicinity of a FR. This issue has been already the subject to many theoretical studies \cite{Ols98,Pet00,Pet01,Yur05,Kim05,Idz06b,Nai07,Pri08} where a zero-range potential approach was used and  much more sophisticated multichannel studies have been performed in Refs.~\cite{Sae08,Mel11,Sae12}. In this paper, we use a finite-range two-channel modeling of the FR resonance. This simple approach permits us to obtain quantitative results in the dimensional reduction issue from the 3D to the low-dimensional behavior, in a broad interval of the colliding energies and of the external magnetic field. Several examples of known resonances illustrate the analysis. The relevance of zero-range approaches in the context of quasi-1D or quasi-2D scattering is studied. 

The paper is organized as follows. In Sec.~\ref{sec:model}, we present the two-channel model used in this paper. We also give the expression of the transition operator of the model which permits us to obtain  straightforwardly  scattering properties in quasi-1D and quasi-2D geometries. In Sec.~\ref{sec:3D}, we consider two  identical atoms without trapping potential. This permits us to recover the mapping between the two-channel model and  the parameters describing the 3D two-body scattering in the low-energy regime. In Sec.~\ref{sec:quasi-2D} (and  \ref{sec:quasi-1D}), we study the two-body scattering  problem in a 2D (and 1D) atomic waveguide. In the monomode regime, the low-dimensional scattering length and effective range parameter in the 2D (and 1D) geometry are exhibited. In Sec.~\ref{sec:bound-states}, we focus on the dimers' properties in confined geometries: their domain of existence, their nature (Feshbach dimers, background dimers, confinement-induced dimers), the role of the confinement, the population of the molecular state. We apply these results to known FR which illustrate three different regimes of collisions (see Table~\eqref{Tab:species}): ${(i)}$ a broad FR of the cesium ($^{133}$Cs) which is characterized also by a large scattering length in the off-resonant regime (vicinity of a shape resonance); ${(ii)}$ two examples of narrow FR with potassium ($^{39}$K) and sodium ($^{23}$Na); and ${(iii)}$ a broad resonance of the lithium ($^7$Li). We compare the binding energies of dimers between the two-channel model and two models of zero-range pseudopotential (the Wigner-Bethe-Peierls model and the effective range model). Our study permits us to give the regimes where the effective range approach is relevant.

\section{Two-channel model}

\label{sec:model}

\subsection{Hamiltonian}

All the two-channel models used in ultra-cold physics capture the heart of the Feshbach mechanism: the scattering resonance is a consequence of the coherent coupling between a pair of atoms in the open channel (i.e. the channel where the scattering process can be observed) and a molecular state in a closed channel \cite{Chi10}. They have been introduced in the context of ultra-cold physics in the study of the BEC-BCS crossover \cite{Hol01}.

The two-channel model of this paper has been used in several studies of the few- and many-body problems. In the few-body problem, it gives quantitative results whereas the equations to be solved are of the same degree of complexity than those obtained with zero-range models \cite{Wer09,Jon10,Mor11a,Pri11c,Tre12}. The model permits us to describe quantitatively the interplay between a shape resonance (where the scattering resonance is due to the direct pairwise interaction between atoms in the open channel) and the FR: a relevant issue in the case of Cs atoms \cite{Mar04,Pri11c}. The resonance width is another important feature of the model, which has important consequences in the few- and many-boson properties \cite{Pet04b,Gog08,Pri13}.

In the model, atoms and molecules are structureless. The pair of atoms in the open channel, characterized by a reduced mass ${\mu}$ and a total mass ${M}$, is coherently coupled  with a  molecular state of mass ${M}$ which belongs to the closed channel. In what follows, ${\hat{H}_0^{\rm a}}$  denotes the free Hamiltonian for two atoms and ${\hat{H}_0^{\rm m}}$  is the free Hamiltonian for a single molecule. Depending on the system under study, the free Hamiltonian may include an external potential acting on the positions of the atoms and of the molecule. In this paper we consider only harmonic external potentials, so that the center of mass and the relative motions are separable. The quantum numbers which label the eigenstates of ${\hat{H}_0^{\rm a}}$  are denoted by ${(\alpha_{\rm c})}$ for the center of mass and ${(\alpha_{\rm r})}$ for the relative particle. The eigenstates of ${\hat{H}_0^{\rm m}}$  for the motion of the molecule are labeled by ${(\alpha_{\rm m})}$:
\begin{align}
&\hat{H}_0^{\rm a} |\alpha_r , \alpha_c \rangle = \left[ \mathcal E_c (\alpha_c) + \mathcal E_r (\alpha_r) \right]  
|\alpha_r,\alpha_c \rangle\\
&\hat{H}_0^{\rm m} |\alpha_m \rangle = \left[ \mathcal E_c (\alpha_m) + E_{\rm mol} \right]  |\alpha_m \rangle ,
\end{align}
where ${E_{\rm mol}}$ is the molecular energy, or internal energy of the molecule. The zero of the energies has been  arbitrarily fixed at the threshold of the atomic continuum in the free space. The molecular energy varies as a function of an applied external magnetic field ${B}$ which permits us to set the scattering length at a desired value. We make the hypothesis that the detuning is sufficiently small that ${E_{\rm mol}}$  can be considered as an affine function of ${B}$. We then introduce the slope in the vicinity of the FR located at ${B=B_0}$:
\begin{equation}
{\delta \mathcal M} = \left( \frac{\partial E_{\rm mol}}{\partial  B}\right)_{B_0}, 
\label{eq:delta_mu}
\end{equation}
which corresponds to the difference between the magnetic moments for an atomic pair in the open channel and the molecular state in the closed channel. The direct interaction between two atoms in the open channel ${\hat{V}_{\rm d}}$ , is modeled by a separable potential \cite{convert}: 
\begin{equation}
\hat{V}_{\rm d} = g |\delta_\epsilon \rangle \langle \delta_\epsilon |  ,
\end{equation}
where ${g}$ is the coupling constant and ${|\delta_\epsilon \rangle}$ is the ket associated with a Gaussian for the relative coordinates of the pair. In the momentum representation:
\begin{equation}
\langle \mathbf k |\delta_\epsilon \rangle \equiv \chi_\epsilon(k) \equiv \exp (-k^2\epsilon^2/4) .
\label{eq:delta_epsilon}
\end{equation}
The parameter ${\epsilon}$ is of the order of the range of the interatomic forces, i.e., of the van der Waals range ${\frac{1}{2} \left( \frac{2 \mu C_6}{\hbar^2} \right)^{1/4}}$ where ${C_6}$ is the London dispersion coefficient. Typically, $\epsilon$ is of the order of few nanometers. In this paper, all the calculations are performed for finite values of ${\epsilon}$. The state ${|\delta_\epsilon\rangle}$ for the relative particle is also used in the coupling operator ${\hat{V}_{\rm c}}$ between a molecular state (${|\Psi^{\rm m}\rangle}$) in the closed channel and a two-atom state (${|\Psi^{\rm a}\rangle}$) in the open channel~\cite{convert}:
\begin{equation}
\langle \Psi^{\rm a} | \hat{V}_{\rm c} |\Psi^{\rm m}\rangle = \Lambda \langle \Psi^{\rm a} | \left( | \delta_\epsilon \rangle  \otimes |\Psi^{\rm m}\rangle \right),
\end{equation}
where ${\Lambda}$ is a real and positive parameter and the center of mass of the pair in the bra  ${\langle \Psi^{\rm a}|}$ acts on the ket ${|\Psi^{\rm m}\rangle}$. Using these notations, the stationary Schr\"{o}dinger equation at energy ${E}$ can be written as
\begin{align}
(E-\hat{H}_0^{\rm a}) |\Psi^{\rm a}\rangle &= \Lambda  | \delta_\epsilon \rangle  \otimes |\Psi^{\rm m}\rangle + g  | \delta_\epsilon \rangle  \otimes 
\langle \delta_\epsilon |\Psi^{\rm a}\rangle ,
\label{eq:sys_p}
\\
(E-\hat{H}_0^{\rm m}) |\Psi^{\rm m}\rangle &= \Lambda  \langle \delta_\epsilon |\Psi^{\rm a}\rangle .
\label{eq:sys_m}
\end{align}

\subsection{Transition operator}

In this section, we derive the transition operator of our two-channel model. As it will be shown, the transition matrix in low dimensions follow straightforwardly from its expression. For this purpose, we introduce the resolvents ${\hat{G}_0^{\rm a}(\lambda)}$, for an atomic pair and ${\hat{G}_0^{\rm m}(\lambda)}$, for a molecule:
\begin{equation}
\hat{G}_0^{\rm a}(\lambda) = (\lambda-\hat{H}_0^{\rm a})^{-1} \quad ; \quad \hat{G}_0^{\rm m}(\lambda) = (\lambda-\hat{H}_0^{\rm m})^{-1} .
\end{equation}
After some algebra, one finds for a scattering process at energy ${E}$ with an incoming state in the open channel ${|\Psi^0\rangle}$:
\begin{multline}
|\Psi^{\rm a}\rangle = |\Psi^0\rangle + \hat{G}_0^{\rm a}(E+i0^+) |\delta_\epsilon \rangle \\
\otimes \left[ g + \Lambda^2  \hat{G}_0^{\rm m}(E) \right] \langle \delta_\epsilon |\Psi^{\rm a}\rangle ,
\end{multline}
where
\begin{multline}
\langle \delta_\epsilon |\Psi^{\rm a}\rangle = \biggl( \mathbb{1}_{\rm mol} - \langle \delta_\epsilon |  \hat{G}_0^{\rm a}(E+i0^+) | \delta_\epsilon \rangle\\
\times \left[ g + \Lambda^2  \hat{G}_0^{\rm m}(E) \right] \biggr)^{-1} \langle \delta_\epsilon|\Psi^0\rangle.
\end{multline}
The incoming state ${|\Psi^0\rangle \equiv |\alpha_r^0,\alpha_c^0\rangle}$ is an eigenstate of ${\hat{H}_0^{\rm a}}$. The form of the coherent coupling in Eqs.~\eqref{eq:sys_p} and Eq.~\eqref{eq:sys_m} implies that the center of mass of the molecule and of the atomic pair have the same quantum numbers: ${\alpha_m^0=\alpha_c^0}$. 

The collisional properties of the atomic pair depend on the relative energy ${E_{\rm rel}}$, i.e., the energy of the pair in its center-of-mass frame:
\begin{equation}
E_{\rm rel} \equiv \mathcal E_r(\alpha_r^0)= E-\mathcal E_c(\alpha_c^0) .
\label{eq:Erel}
\end{equation}
We continue our study in the center of mass frame, where the wave function for the relative motion of the pair at energy ${E_{\rm rel}}$ is denoted by ${|\Psi_{\rm rel}\rangle}$. The general expression for the scattering state is
\begin{equation}
|\Psi_{\rm rel} \rangle = |\alpha_r^0 \rangle + \hat{G}_0^{\rm rel}(E_{\rm rel} +i0^+) \hat{T}(E_{\rm rel}+i0^+) |\alpha_r^0 \rangle,
\label{eq:scatt_rel}
\end{equation}
where the transition operator of the two-channel model is
\begin{equation}
\hat{T}(\lambda) = \frac{| \delta_\epsilon  \rangle \langle \delta_\epsilon|}
{ \left(g + \frac{\Lambda^2}{\lambda-E_{\rm mol}}\right)^{-1} 
- \langle \delta_\epsilon |  \hat{G}_0^{\rm rel}(\lambda) | \delta_\epsilon \rangle} ,
\label{eq:T_operator}
\end{equation}
and ${\hat{G}_0^{\rm rel}(\lambda)}$ is the resolvent for the relative particle which depends on the external potential:
\begin{equation}
\hat{G}_0^{\rm rel}(\lambda)= \left(\lambda-\frac{\hat{p}^2}{2\mu}-\hat{V}_{\rm trap}^{\rm rel}\right)^{-1} .
\label{eq:Green_rel}
\end{equation}
In Eq.~\eqref{eq:Green_rel}, ${\hat{p}}$ is the relative momentum of the pair and ${\hat{V}_{\rm trap}^{\rm rel}}$ is the part of the trap potential which acts only on the relative particle. 

\section{Two particles in the free space}

\label{sec:3D}
In this section, we recall the mapping between the parameters of the model (${E_{\rm mol}, {\delta \mathcal M}, \epsilon, \Lambda, g}$) and the low-energy scattering features for two particles in the free space. In this situation, there is no external potential and ${\alpha_r = \mathbf k}$ is the momentum of the relative particle. 

The on-shell transition matrix for a scattering process at a relative energy ${\lambda=E_{\rm rel}>0}$ is deduced from Eq.~\eqref{eq:T_operator} which gives the transition matrix of Refs.~\cite{Jon10,Pri11c,convert}. The standard expression for the scattering length as a function of the external magnetic field in Eq.~\eqref{eq:a} is obtained by the mapping
\begin{align}
&{{ B}-{ B}_0} = \frac{E_{\rm mol}}{{\delta \mathcal M}} - \frac{\Lambda^2}{g {\delta \mathcal M}} + {\Delta  B} \label{eq:detuning}, \\
&\Delta { B} =  \frac{2 \pi \hbar^2 \Lambda^2 a_{\rm bg}}{\mu g^2 {\delta \mathcal M}} \label{eq:DeltaB} , \\
&a_{\rm bg} = \sqrt{\frac{\pi}{2}} \frac{\epsilon g}{g-g^{\rm c}} \quad  \mbox{with}\quad 
g^{\rm c} = - \frac{\sqrt{2}\pi^{3/2}\hbar^2\epsilon}{\mu} \label{eq:abg} .
\end{align}
In this paper, we also use the width parameter $R^\star$ \cite{Pet04b} 
\begin{equation}
R^\star = \frac{\hbar^2}{2\mu a_{\rm bg} \delta \mathcal M \Delta B}.
\label{eq:Rstar}
\end{equation}
This length, which is always positive, characterizes the width of resonance: for narrow resonances, it is large as compared to the range of the interatomic forces. 

Using these definitions, for an incoming wave number ${k_{\rm rel}}$ and an energy ${E_{\rm rel}=\hbar^2 k_{\rm rel}^2/(2\mu)}$, one can introduce an energy dependent scattering length  
\begin{equation}
a_{\rm eff}(E_{\rm rel}) = [\chi_\epsilon(k_{\rm rel})]^2  
\left[\frac{1}{a} + \frac{R^\star(1-\frac{a_{\rm bg}}{a})^2k_{\rm rel}^2}
{R^\star a_{\rm bg}(1-\frac{a_{\rm bg}}{a})k_{\rm rel}^2+1}\right]^{-1} .
\label{eq:a_eff}
\end{equation}
The length ${a_{\rm eff}}$ permits us to simplify the expression of the on-shell transition matrix in the free space with:
\begin{equation}
\label{eq:Tmatrix_3D}
\langle \mathbf k_{\rm rel} | \hat{T}(E_{\rm rel}) |\mathbf k_{\rm rel} \rangle 
= \frac{\frac{2\pi\hbar^2}{\mu}}
{\frac{1}{a_{\rm eff}(E_{\rm rel})} 
+ i k_{\rm rel} {\rm erfc} \left(\frac{-i\epsilon k_{\rm rel}}{\sqrt{2}}\right)} .
\end{equation}
We close this section by recalling the main results on the 3D bound states of the model. At this stage, it is worth pointing out that in this paper, a dimer denotes a bound state of two particles in the open channel and is not a molecule which belongs to the closed channel. The binding energies of the dimers ${E_{\rm dim}= \frac{-\hbar^2q^2}{2\mu}}$, where the binding wave number ${q}$ is positive  (${q>0}$), are given by the poles of the transition matrix in Eq.~\eqref{eq:Tmatrix_3D}. They have been already studied in Ref.~\cite{Pri11c}. One can distinguish two types of dimers: ${(i)}$ the background dimer (BD), which results from the direct interaction in the open channel without interchannel coupling; ${(ii)}$ the Feshbach dimer (FD), which results from the coupling between a pair of atoms and the molecular state. The Feshbach dimer can be seen as a contamination of the open channel by the molecular state due to the coupling with pairs of atoms. For a large and negative energy detuning, the binding energy of the Feshbach dimer verifies
\begin{equation}
E_{\rm dim} = E_{\rm mol}+\frac{\delta \mathcal M \Delta B}{E_{\rm mol}}
\frac{\hbar^2}{2\mu\epsilon^2}\frac{\sqrt{\frac{2}{\pi}}\frac{a_{\rm bg}}{\epsilon}}{(1-\sqrt{\frac{2}{\pi}}\frac{a_{\rm bg}}{\epsilon})^2} + \dots
\label{eq:Edim_large_detuning}
\end{equation}
As expected, it tends to the molecular energy, meaning that the occupation probability of the pair in the open channel tends to zero. 

In this model, the BD exists only in the domain where ${a_{\rm bg} > \sqrt{\pi}\epsilon/\sqrt{2}}$ and the FD exists for negative-energy detuning, \textit{i.e.} for ${\delta \mathcal M (B-B_0)<0}$. In the presence of a BD, in the intermediate regime of negative energy detuning there is a level crossing between the BD and the FD corresponding to a large hybridization between the two types of dimers. Consequently, the denomination of BD and FD is somehow arbitrary in this region. The domain of existence of the dimers is summarized in~Table~\eqref{tab:dimers_3D}.
\begin{table}
\begin{tabular}{ccc}
\hline
\hline
 & ${\qquad \delta \mathcal M (B-B_0) <0 \quad }$ & ${\quad  \delta \mathcal M (B-B_0) >0}$ \\
\hline
${a_{\rm bg} < \epsilon \sqrt{\frac{\pi}{2}}}$ & FD &  no dimer   \\
${a_{\rm bg} > \epsilon \sqrt{\frac{\pi}{2}}}$ & FD + BD  & BD \\
\hline
\hline
\end{tabular}
 \caption{Domain of existence of the Feshbach dimer (denoted FD) and background dimer (denoted BD) in the free space as a function of the energy detuning ${{\delta \mathcal M} \times  ({B}-{B}_0)}$.}
\label{tab:dimers_3D}
\end{table}

In Table~\eqref{Tab:species}, we have gathered the values of the parameters ${a_{\rm bg}, \epsilon, R^\star}$ for FR taken as examples in this paper. The parameter ${\epsilon}$ is of the order of the range of the interatomic forces. When spectroscopic data or \textit{ab initio} predictions on dimers binding energies are available, a precise value of the short-range parameter ${\epsilon}$ can be determined. In this case, the model permits us to describe quantitatively the dimer spectrum for large detunings \cite{Jon10,Pri11c}. In other cases,  we have set ${\epsilon}$ at the value of the van der Waals range.
\begin{table}[hx]
\begin{tabular}{cccccc}
\hline
\hline
Species $~~$ & $B_0$~(G) $~~$ & $\Delta B$~(G) $~~$ & $a_{\rm bg}/a_0$ $~~$ & $\epsilon/a_0$ $~~$ & $R^\star/a_0$  \\
\hline
$^{133}$Cs & ${-11.7}$ & ${28.7}$ & ${1720}$ &${100}$& ${0.13}$\\
$^{39}$K & ${752.4}$ & $-0.4$  & $-35$ & ${65}$& ${1200}$\\
$^7$Li & ${736.8}$ & ${-192.3}$ & ${-25}$ & $33$& ${39.6}$\\
$^{23}$Na & ${853}$ & ${0.0025}$ & $63$ & $44$& ${2.1 \times 10^5}$\\
\hline
\hline
\end{tabular}
\caption{Characteristic lengths of FR considered in this paper \cite{Chi10,Der07}. The length ${a_0}$ is the Bohr radius, ${B_0}$~is the magnetic field value at resonance, ${a_{\rm bg}}$ is the background scattering length and ${R^\star}$ is defined in Eq.~\eqref{eq:Rstar}.}
\label{Tab:species}
\end{table}

\section{Quasi-2D geometry}

\label{sec:quasi-2D}

\subsection{Scattering in a 1D transverse harmonic trap}

We consider a planar atomic waveguide consisting of a harmonic trap of atomic frequency ${\omega_\perp}$ in the $x$ direction. The particles are thus free to move along the two other directions. The characteristic length of the trap for the relative particle is denoted by
\begin{equation}
a_\perp = \sqrt{\frac{\hbar}{\mu \omega_\perp}} .
\label{eq:def_aperp}
\end{equation}
The quantum numbers for the relative motion are denoted by ${\alpha_r\equiv(\mathbf k,n)}$, where ${\mathbf k}$ is a 2D vector and ${n}$ labels the states of the harmonic oscillator. The corresponding eigen-energies are
\begin{equation}
\mathcal E_r(\mathbf k,n)=\frac{\hbar^2k^2}{2\mu} + \hbar \omega_\perp \left(n +\frac{1}{2}\right) .
\end{equation}
In what follows, ${\phi_n(x)}$ denotes the eigenfunctions of the harmonic trap: 
\begin{equation}
\phi_n(x) \equiv \langle x | n \rangle = \frac{1}{\pi^{1/4}} 
\frac{\exp(-\frac{x^2}{2a_\perp^2})}{\sqrt{a_\perp 2^n n!}}  H_n(x/a_\perp) .
\end{equation}
For convenience, we use in this paper the following reduced parameters:
\begin{equation}
\tau \equiv \frac{\lambda}{2\hbar \omega_\perp}-\frac{1}{2D} \quad \text{and} \quad \eta \equiv \frac{\epsilon^2}{2a_\perp^2} .
\label{eq:def_tau}
\end{equation}
For the planar atomic waveguide, ${D=2}$ and thus ${\tau={\lambda}/{2\hbar \omega_\perp}-{1}/{4}}$.

We now consider a scattering process where the incoming wave of relative energy ${E_{\rm rel}=E_0}$ ${\left(E_0 \ge \frac{\hbar\omega_\perp}{2}\right)}$ populates only one transverse mode ${n_0}$. The interaction acts only in the ${s}$-wave sector, consequently, scattering occurs iff ${n_0}$ is even and the outcoming state can populate only even transverse modes. We then introduce the integer ${p_0=n_0/2}$. The incoming state is characterized by a 2D momentum ${\mathbf k_0}$ [i.e., ${\alpha_r^0=(\mathbf k_0, 2p_0)}$] and ${E_0=\mathcal E_r(\mathbf k_0, 2p_0)}$ with
\begin{equation}
E_0= \frac{\hbar^2 k_0^2}{2\mu} +\hbar\omega_\perp \left( 2 p_0 + \frac{1}{2}\right) 
\quad ; \quad 
\tau_0 =  \frac{E_0 - \frac{\hbar \omega_\perp}{2}}{2\hbar \omega_\perp} . 
\label{eq:def_tau0_2D}
\end{equation}
The outcoming state is a coherent superposition of cylindrical waves characterized by the quantum numbers
${\alpha_r=(\mathbf q_p, 2p)}$. The conservation of energy gives
\begin{equation}
q_p = \sqrt{\frac{2\mu E_0}{\hbar^2} - \frac{4 p + 1 }{a_\perp^2}} .
\label{eq:k_p}
\end{equation}
Equation~\eqref{eq:k_p} shows that after the scattering process, for a state of relative energy ${E_0>\hbar\omega_\perp/2}$, only the transverse modes of quantum numbers smaller or equal to ${2 \tau_0}$ are populated in the open channel. Hence, if ${\frac{\hbar\omega_\perp}{2}<E_0<\frac{5\hbar\omega_\perp}{2}}$, only the first transverse mode is populated for large interatomic distances ${\rho}$ ${(\rho \gg \epsilon, a_\perp)}$: this corresponds to the monomode regime of the planar atomic waveguide. For ${E_0\ge \frac{5\hbar\omega_\perp}{2}}$, the number of even transverse modes populated after the scattering process is given by ${\lfloor \tau_0 \rfloor}$ where ${\lfloor  \cdot  \rfloor}$ indicates the integer part. For increasing values of the relative energy ${E_0}$, a new 2D-continuum is opened when ${\tau_0}$ crosses an integer value. 

Equation~\eqref{eq:T_operator} shows that the scattering properties in the planar atomic waveguide are deduced from the behavior of the function 
\begin{equation}
\langle \delta_\epsilon | \hat{G}_0^{\rm rel}(\lambda) | \delta_\epsilon \rangle 
= \int \frac{d^2k}{(2\pi)^2} \sum_{n=0}^\infty 
\frac{|\chi_\epsilon(k) \langle \delta_\epsilon^{\rm 1D} | n \rangle|^2}
{\lambda-\mathcal E_r(\mathbf k,n)} .
\label{eq:I_2D}
\end{equation}
In Eq.~\eqref{eq:I_2D}, ${|\delta_\epsilon^{\rm 1D}\rangle}$ is the Gaussian state associated with the transverse degree of freedom. In the configuration space, its representation is 
\begin{equation}
\langle x |\delta_\epsilon^{\rm 1D} \rangle = \frac{1}{\epsilon \sqrt{\pi}}\exp(-x^2/\epsilon^2) .
\label{eq:delta_epsilon_x}
\end{equation}
The transition matrix between the incoming state ${|\alpha_r^0 \rangle=| \mathbf k_0, 2p_0 \rangle}$ and the outcoming state ${|\alpha_r \rangle = |\mathbf q_p, 2p\rangle}$ can be written:
\begin{multline}
\langle \alpha_r |\hat{T}(E_0+i0^+)|\mathbf \alpha_r^0\rangle
= 
\left(\frac{1-\eta}{1+\eta}\right)^{p_0+p} \frac{\phi_{2p}(0) \phi_{2p_0}(0)}{|\phi_0(0)|^2}
\\
\times \frac{2 \pi\hbar^2}{ \mu}
\frac{e^{\frac{(k_0 ^2-q_p^2)\epsilon^2}{4}} e^{4\eta p_0}}
{\frac{\sqrt{\pi} (1+\eta) a_\perp e^{-\eta}}{a_{\rm eff}(E_0)} + J_2(\eta,\tau_0+i0^+)} ,
\label{eq:transition_matrix_2D}
\end{multline}
where the function ${J_2(\eta,\tau)}$ is defined by:
\begin{multline}
J_2(\eta,\tau) \equiv -\frac{2\pi\hbar^2}{\mu}  \frac{e^{4\eta \tau}}
{|\langle \delta_\epsilon^{\rm 1D} | n=0 \rangle|^2}\\
\times \left[ 
\langle \delta_\epsilon | \hat{G}_0^{\rm rel}(\lambda) | \delta_\epsilon \rangle
+\frac{\mu}{\sqrt{2}\pi^{3/2}\hbar^2\epsilon} \right] ,
\end{multline}
where the energy ${\lambda}$ is given by Eq.~\eqref{eq:def_tau}. In the configuration space and in the limit of large interparticle distances, the scattering state is
\begin{multline}
\langle \boldsymbol{\rho},x|\Psi_{\rm rel}^{\mathbf k_0, 2p_0}\rangle \underset{\rho\gg \epsilon,a_\perp}{=} 
\phi_{2p_0}(x) e^{i\mathbf k_0 \cdot \boldsymbol{\rho}}
-  \frac{\mu}{\hbar^2} \sum_{p=0}^{\lfloor \tau_0 \rfloor} \phi_{2p}(x)\\
\times
\frac{e^{i\left(q_p \rho +\frac{\pi}{4}\right)}}{\sqrt{2\pi q_p \rho}} 
\langle q_p \hat{e}_{\boldsymbol \rho} ,2p |\hat{T}(E_0+i0^+)| \mathbf k_0
, 2p_0 \rangle .
\label{eq:2D_scatt_large_rho_sur_epsilon}
\end{multline}

In what follows, we study the behavior of the function ${J_2(\eta,\tau)}$ which is in the denominator of the $T$-matrix in Eq.~\eqref{eq:transition_matrix_2D}. For this purpose, we perform the change of variable ${y=k a_\perp/2}$ in Eq.~\eqref{eq:I_2D} and consider the domain of negative reduced energies: ${\tau<0}$. This permits us to use the transformation
\begin{equation}
\frac{1}{\tau-y^2-n/2} = - \int_0^\infty du \, e^{u(\tau -n/2-y^2)} .
\end{equation}
The summation over the variable ${n}$ in Eq.~\eqref{eq:I_2D} is achieved in the Appendix (\ref{ap:Thesum}). After integration over the variable ${y}$, one obtains:
\begin{equation}
J_2(\eta,\tau) = \int_0^\infty \frac{du}{u+4\eta}\frac{e^{\tau (u+4\eta)}}
{\sqrt{1 -e^{-u} \left(\frac{1-\eta}{1+\eta}\right)^2}} - \frac{(1+\eta)e^{4\eta\tau}}{\sqrt{\eta}}.
\label{eq:J_eta_integral_2D}
\end{equation}
In the limit of a vanishingly small reduced energy ${(\tau \to 0)}$, the function ${J_2(\eta,\tau)}$ exhibits a logarithmic singularity, which is a characteristic of the quasi-2D geometry: 
\begin{equation}
J_2(\eta,\tau) = - \ln \left(-\tau e^\gamma \right) - \mathcal C^{\rm 2D}_0(\eta) + \mathcal C^{\rm 2D}_1(\eta) \tau + O(\tau^2) .
\label{eq:J_small_tau_2D}
\end{equation}
In Eq.~\eqref{eq:J_small_tau_2D},  ${\gamma=0.577\dots}$ is the Euler constant,
\begin{multline}
\mathcal C^{\rm 2D}_0(\eta) =  \sqrt{\eta} - \int_0^\infty \frac{du}{u+4\eta} \\
\times
\left[
\frac{1}{\sqrt{1-e^{-u}\left(\frac{1-\eta}{1+\eta}\right)^2 }}-\frac{1}{\sqrt{u+4\eta}}
-\frac{u+4\eta}{u+1}
\right]
\label{eq:C2D_0}
\end{multline} 
and 
\begin{equation}
\mathcal C^{\rm 2D}_1(\eta) = 2\ln \left[ \frac{2(1+\eta)}{(1+\sqrt{\eta})^2}\right]
-4 \eta -4 (1+\eta)\sqrt{\eta} .
\label{eq:C2D_1}
\end{equation}
In the general case where ${\tau}$ is a complex variable on the physical sheet (i.e., the complex plane with the branch cut ${\mathbb R^{+}}$), the function ${J_2(\eta,\tau)}$ can be represented by the following series expansion:
\begin{multline}
J_2(\eta,\tau) = -\frac{(1+\eta)}{\sqrt{\eta}} \left(e^{4\eta\tau}-1\right) - \mathcal C^{\rm 2D}_0(\eta) + \ln (4\eta) \\
+ E_1(-4\eta\tau)  - \sum_{p=1}^{\infty} \left(\frac{1-\eta}{1+\eta}\right)^{2p} \frac{(2p-1)!!}{(2p)!!} \\
\times e^{4\eta p} \biggl\{E_1(4 \eta p) - E_1[4\eta(p-\tau)]\biggr\}.
\label{eq:J_series_2D}
\end{multline}
When a continuum is opened in the scattering process, one uses the analytic continuation for the exponential integral function ${E_1}$ in the calculation of ${J_2(\eta,\tau_0+i0^+)}$  defined in Eq.~\eqref{eq:J_series_2D} \cite{abramowitz}:  
\begin{equation}
E_1(-x-i0^+) = - E_i(x) + i\pi \quad (x>0) .
\end{equation}

\subsection{Low energy and zero-range limit}

\label{sec:2D_relevance_of_ZR}
In usual situations, the typical radius of interatomic forces is negligible as compared to the transverse length ${a_\perp}$ (${\epsilon \ll a_\perp}$ or equivalently ${\eta \ll 1}$). Thus, in Eq.~\eqref{eq:J_small_tau_2D} one can use the zero-range limit of the parameters defined in Eqs.~\eqref{eq:C2D_0} and~\eqref{eq:C2D_1}:
\begin{align}
&\lim_{\eta \to 0} \mathcal C^{\rm 2D}_0(\eta) =   \int_0^\infty \frac{du}{u} \biggl[
\frac{-1}{\sqrt{1-e^{-u}}}+\frac{1}{\sqrt{u}} +\frac{u}{u+1} \biggr] \\
& \phantom{\lim_{\eta \to 0} \mathcal C^{\rm 2D}_0(\eta)}  = 1.3605 \dots , \\
&\lim_{\eta \to 0} \mathcal C^{\rm 2D}_1(\eta) = 2\ln (2).
\end{align}
We now consider a low-energy process. At large interatomic distances, only few transverse modes are populated, and the  product ${\eta \tau_0}$ is very small as compared to unity. One can then use with a high accuracy the zero-range limit for ${J_2(\eta,\tau)}$ by using the integral representation
\begin{equation}
\lim_{\eta \to 0} J_2(\eta,\tau)= {\rm P}_f \int_0^\infty \frac{du}{u}\frac{e^{\tau u}}
{\sqrt{1-e^{-u}} },
\end{equation}
where ${{\rm P}_f}$ denotes the Hadamard finite part of the integral \cite{Sch78}. In this regime, one can also perform the zero-range limit in the series expansion of Eq.~\eqref{eq:J_series_2D}:
\begin{multline}
\lim_{\eta \to 0} J_2(\eta,\tau_0+i0^+) =  \sum_{p=1}^{\infty} \frac{(2p-1)!!}{(2p)!!}
 \ln\left|\frac{p}{p-\tau_0} \right|\\
 +\ln \left( \frac{-B}{2\pi \tau_0 }\right) +i \pi\left[ \frac{(2 \lfloor \tau_0 \rfloor +1)!!}{(2 \lfloor \tau_0 \rfloor)!!} -1 \right]
\label{eq:J_series_2D_ZRP}
\end{multline} 
where ${B=2\pi\times \exp\left[-\gamma- C^{\rm 2D}_0(0)\right]=0.9049\dots}$. Equation \eqref{eq:J_series_2D_ZRP} permits to recover the results  deduced by using the zero-range potential approach \cite{Pet01,Idz06b,Nai07,Pri08}.

\subsection{Monomode regime in a 2D atomic waveguide}

\subsubsection{Scattering amplitude and resonance}

In the monomode regime, the collisional energy $E_0$ verifies ${\frac{\hbar\omega_\perp}{2} <E_0 <\frac{5\hbar\omega_\perp}{2}}$.  The 2D scattering amplitude ${f_{\rm 2D}}$ is defined from the behavior of the wave function at large distances:
\begin{equation}
\langle \boldsymbol{\rho},x|\Psi_{\rm rel}^{\mathbf k_0 ,0}\rangle \sim
 \phi_{0}(x) \left[
e^{i\mathbf k_0 \cdot \boldsymbol{\rho}}
+  \frac{\sqrt{\pi} f_{\rm 2D}(k_0) }{\sqrt{2k_0 \rho}} {e^{i \left(k_0 \rho +\frac{\pi}{4}\right)}}
  \right] .
\label{eq:2D_scatt_lowE} 
\end{equation}
The 2D scattering amplitude ${f_{\rm 2D}}$ is the basic ingredient which characterizes the two-body interaction in 2D dilute systems. Near the threshold of the propagating regime ${\left(E_0 \to \frac{\hbar\omega_\perp}{2}^+\right)}$, in the effective range approximation, the 2D scattering amplitude is of the form
\begin{equation}
f_{\rm 2D}(k_0)=\frac{1}
{\ln(k_0 a_{\rm 2D} e^\gamma/2) - a_{\rm 2D} R_{\rm 2D} k_0^2 - \frac{i\pi}{2}} .
\label{eq:f2D_effective_range}
\end{equation}
In Eq.~\eqref{eq:f2D_effective_range}, we have introduced the 2D scattering length
\begin{equation}
a_{\rm 2D} = a_\perp \exp\left[\frac{\mathcal C^{\rm 2D}_0(\eta)-\gamma}{2} -\frac{\sqrt{\pi} (1+\eta) a_\perp e^{-\eta}}{2 a_{\rm eff}(\hbar\omega_\perp/2)}\right] ,
\label{eq:a2D}
\end{equation}
and the 2D effective range parameter ${R_{\rm 2D}}$:
\begin{multline}
R_{\rm 2D} =  \mathcal C^{\rm 2D}_1(\eta) \frac{a_\perp^2}{8 a_{\rm 2D}} + \frac{\sqrt{\pi}}{2} (1+\eta) \frac{a_\perp}{a_{\rm 2D}} \biggl[\frac{\epsilon^2e^{-\eta}}{2a_{\rm eff}(\hbar\omega_\perp/2)} 
\\ 
+ \frac{a_\perp^4 }{R^\star a_{\rm bg}^2} \left(\frac{{a}/{a_{\rm bg}}-1}{{a}/{a_2^{\rm div}-1}} \right)^2   
\biggr] .
\label{eq:R2D}
\end{multline}
The length ${a_2^{\rm div}}$ which appears in Eq.~\eqref{eq:R2D} corresponds to the value of the 3D scattering length ${a}$, such that ${a_{\rm 2D}}$ diverges;  more precisely, ${a_{\rm 2D} \to + \infty}$ for ${a \to (a_2^{\rm div})^-}$. The expression of ${a_2^{\rm div}}$ is given by 
\begin{equation}
a_2^{\rm div} = \frac{a_{\rm bg}}{1+ {2 \delta \mathcal M \Delta B}/{\hbar \omega_\perp}} .
\label{eq:resonance_2D_b}
\end{equation}
The 2D scattering length ${a_{\rm 2D}}$ in Eq.~\eqref{eq:a2D} can be tuned from ${0}$ to ${+\infty}$ by playing with the value of the external magnetic field. The resonant regime in 2D scattering occurs for large values of ${a_{\rm 2D}/a_\perp}$, that is, when the 3D scattering length $a$ is smaller and almost equals to
to ${a_2^{\rm div}}$. The resonance corresponds to a maximum in the scattering amplitude which appears at a 2D collisional momentum ${k_0 \sim k_{\rm res} = \frac{2e^{-\gamma}}{a_{\rm 2D}}}$. The width of this scattering resonance ${\Delta k/k_0}$ is of the order of unity. 

\subsubsection{2D zero-range models}

\label{ap:zero-range}

The expression of the 2D scattering amplitude in Eq.~\eqref{eq:f2D_effective_range} can be  deduced exactly from a purely 2D and zero-range potential model that we denote as the 2D effective range model (2D-ERM). In this approach, the two-body wave function for the relative motion denoted by ${\psi_{\rm 2D}}$, depends only on the 2D relative coordinates ${\boldsymbol \rho}$ . The interaction is replaced by imposing the following contact condition on the wave function:
\begin{multline}
\lim_{\rho \to 0} \biggl[ 1 -\ln\left(\frac{\rho}{  a_{\rm 2D} }\right) \rho \partial_\rho \\
+ a_{\rm 2D} R_{\rm 2D} \rho \partial_\rho \Delta_\rho \biggr] 
\langle \boldsymbol{\rho} | \hat{\Pi}_s^{\rm 2D} | \psi_{\rm 2D} \rangle = 0,
\label{eq:contact_2D}
\end{multline}
where  ${\hat{\Pi}_s^{\rm 2D}}$ is the projector on the 2D $s$-wave channel:
\begin{equation}
\langle \boldsymbol{\rho} | \hat{\Pi}_s^{\rm 2D} | \psi_{\rm 2D} \rangle = \int \frac{d\theta}{2\pi}
\langle \boldsymbol{\rho} |  \psi_{\rm 2D} \rangle
\,\,\, \text{where} \,\, {\theta= \angle (\boldsymbol \rho, \hat{\bf e}_y)}.
\label{eq:projector_2D}
\end{equation}
In the few- and many-body problem, the contact condition is imposed on the wave function for each interacting pair of atoms and the limit on the relative coordinates is taken for a fixed value of the other coordinates. Equation \eqref{eq:contact_2D} generalizes the 2D Wigner-Bethe-Peierls (WBP) model where the effective range term is set to zero  ${(R_{\rm 2D}=0)}$. 

\section{Quasi-1D geometry}

\label{sec:quasi-1D}

\subsection{Scattering in a 2D transverse harmonic trap}

In this section, the pair of atoms move in a linear atomic waveguide made of a harmonic trap with the same atomic frequency ${\omega_\perp}$ in the $x$ and $y$ directions.  The quantum numbers for the relative motion are denoted by ${\alpha_r\equiv(k,n,m)}$ where ${m\hbar}$ is the angular momentum along $z$ and 
\begin{equation}
\mathcal E_r(k,n,m)=\frac{\hbar^2k^2}{2\mu} + \hbar \omega_\perp (2n+|m|+1) .
\end{equation}
The eigenfunctions of the transverse harmonic oscillator are given by
\begin{multline}
\langle \mathbf  r_\perp|n,m\rangle = \left[\frac{\pi a_\perp^2(n+|m|)!}{n!} \right]^{-1/2} \left(\frac{r_\perp}{a_\perp}\right)^{|m|} e^{im\theta}\\
\times e^{-\frac{1}{2}(r_\perp/a_\perp)^2}L_n^{(|m|)}(r_\perp^2/a_\perp^2) .
\label{eq:fo_OH_2D}
\end{multline}
In Eq.~\eqref{eq:fo_OH_2D},  ${L_n^{(\alpha)}}$ is the generalized Laguerre polynomial and ${ \mathbf  r_\perp = (r_\perp,\theta)}$ are the polar coordinates in the ${x-y}$ plane.

In this section, we use the reduced energy defined in Eq.~\eqref{eq:def_tau} with ${D=1}$ that is ${\tau=\lambda/(2\hbar \omega_\perp)-1/2}$. For a scattering state of incoming wave ${\alpha_r^0\equiv(k_0,n_0,m_0)}$, the reduced energy ${\tau=\tau_0}$ is 
\begin{equation}
\tau_0 = \frac{k_0^2 a_\perp^2}{4} + n_{0} + \frac{|m_{0}|}{2} .
\end{equation}
Similarly to the 2D atomic waveguide, the threshold of each 1D-continuum corresponds to an integer value of ${\tau_0}$. The integer ${\lfloor \tau_0 \rfloor}$ is the number of  occupied transverse states in the outcoming state. Due to the $s$-wave character of the interaction, the matrix elements are non zero only between eigenstates with zero angular momentum. For each transverse state ${|n,m=0\rangle}$ occupied by the outcoming wave, we define the wave number ${q_n}$, obtained by energy conservation:
\begin{equation}
q_n = \sqrt{\frac{2\mu E_0}{\hbar^2} - \frac{4n+2}{a_\perp^2}} .
\label{eq:1D_k_p}
\end{equation}
Using this notation, the $T$-matrix can be written as
\begin{multline}
\langle q_n, n, m |\hat{T}(E_0+i0^+)|k_0, n_0, m_0 \rangle = \delta_{m,0}\delta_{m_0,0} \frac{2 \pi \hbar^2}{\mu \sqrt{\pi}a_\perp} \\
\times  \left(\frac{1-\eta}{1+\eta}\right)^{n+n_0}
\frac{ e^{\frac{(k_0^2-q_n^2)\epsilon^2}{4}} e^{4\eta n_0}}
{\frac{ \sqrt{\pi}a_\perp (1+\eta)^2  e^{-2\eta}}{a_{\rm eff}(E_0) } +  J_1(\eta,\tau_0+i0^+)} ,
\label{eq:transition_matrix_1D}
\end{multline}
where the function ${J_1(\eta,\tau)}$ is defined by
\begin{multline}
J_1(\eta,\tau) \equiv -\frac{2\pi\hbar^2}{\mu} \frac{|\phi_0(0)|^2 e^{4\eta\tau}}
{|\langle \delta_\epsilon^{\rm 1D} | n=0 \rangle|^4}
\\
\times 
\left[
\langle \delta_\epsilon | \hat{G}_0^{\rm rel}(\lambda) | \delta_\epsilon \rangle
+
\frac{\mu}{\sqrt{2}\pi^{3/2}\hbar^2\epsilon}
\right] ,
\label{eq:def_J1}
\end{multline}
where the energy ${\lambda}$ is given by Eq.~\eqref{eq:def_tau}. In the limit of large interparticle distances (${z \gg \epsilon, a_\perp}$), the scattering wave function verifies:
\begin{multline}
\langle z,\mathbf r_\perp|\Psi_{\rm rel}^{k_0, n_0,0}\rangle \underset{z\gg \epsilon,a_\perp}{=} 
\langle \mathbf r_\perp | n_0,0\rangle e^{i k_0 \cdot z} 
-  \frac{i \mu}{\hbar^2} \sum_{n=0}^{\lfloor \tau_0 \rfloor}  \frac{e^{iq_n|z|}}{q_n} \\
\times \langle \mathbf r_\perp | n,0\rangle
 \langle q_n,n,0|\hat{T}(E_0 +i0^+)| k_0, n_0,0 \rangle .
\label{eq:1D_scatt_large_rho_sur_epsilon}
\end{multline}

Following the same reasoning as in Sec.~\eqref{sec:quasi-2D}, one obtains a simple expression of ${J_1(\eta,\tau)}$ for negative values of ${\tau}$: 
\begin{equation}
J_1(\eta,\tau) = \int_0^\infty \frac{du}{\sqrt{u+4\eta}}\frac{e^{\tau (u+4\eta)}}
{1-e^{-u} \left(\frac{1-\eta}{1+\eta}\right)^2} - \frac{(1+\eta)^2e^{4\eta\tau}}{\sqrt{\eta}}.
\label{eq:J_1_integral}
\end{equation}
In the limit of vanishingly small reduced energy ${\tau \to 0}$, the function ${J_1(\eta,\tau)}$ has the following behavior: 
\begin{equation}
J_1(\eta,\tau) = \sqrt{\frac{\pi}{-\tau}}+\mathcal C^{\rm 1D}_0(\eta) + \mathcal C^{\rm 1D}_1(\eta) \tau +  O(\tau^2),
\label{eq:J_1_small_tau}
\end{equation}
where
\begin{multline}
\mathcal C^{\rm 1D}_0(\eta) =  \int_0^\infty \frac{du}{\sqrt{u+4\eta}} \biggl[
\frac{1}{\left(\frac{1+\eta}{1-\eta}\right)^2 e^u-1}-\frac{(1+\eta)^2}{4\eta+u} \biggr] \\
- 4\sqrt{\eta} 
\end{multline}
and
\begin{equation}
\mathcal C^{\rm 1D}_1(\eta) =  -\frac{16}{3} \eta^{\frac{3}{2}}- 4 (1+\eta)^2\sqrt{\eta} +\int_0^\infty du \frac{\sqrt{u+4\eta}}{\left(\frac{1+\eta}{1-\eta}\right)^2 e^u-1} .
\end{equation}
When the reduced energy is a complex variable in the physical sheet, one can use a series representation for the function ${J_1(\eta,\tau)}$:
\begin{multline}
J_1(\eta,\tau) = \sqrt{\frac{\pi}{-\tau}} {\rm erfc} \left(2 \sqrt{-\eta\tau} \right) +4\sqrt{\eta}   
+ \mathcal C^{\rm 1D}_0(\eta) \\- \frac{(1+\eta)^2}{\sqrt{\eta}} \left[e^{4\eta\tau} -1 \right] +  
\sqrt{\pi} \sum_{p=1}^{\infty}  e^{4\eta p} \left( \frac{1-\eta}{1+\eta}\right)^{2p}\\
\times \left[ \frac{{\rm erfc}\left(2\sqrt{\eta(p-\tau)} \right)}{\sqrt{p-\tau}} - 
\frac{{\rm erfc}\left(2\sqrt{\eta p} \right)}{\sqrt{p}}\right] .
\label{eq:J_series}
\end{multline}

\subsection{Low energy scattering and zero-range limit}

As was done for the 2D atomic waveguide in Sec.~\eqref{sec:2D_relevance_of_ZR}, we consider the relevance of the zero-range limit. For usual trapping frequencies, the typical radius of interatomic forces is negligible as compared to the transverse length ${a_\perp}$ (${\epsilon \ll a_\perp}$ or equivalently ${\eta \ll 1}$). Thus, one can use the zero-range limits of the parameters in Eq.~\eqref{eq:J_1_small_tau}:
\begin{equation}
\lim_{\eta \to 0} \mathcal C^{\rm 1D}_0(\eta) =  \sqrt{\pi} \zeta(1/2)
\ ; \
\lim_{\eta \to 0} \mathcal C^{\rm 1D}_1(\eta) = \frac{\sqrt{\pi}}{2} \zeta(3/2),
\end{equation}
where ${\zeta}$ is the Riemann zeta function. 

In the low-energy regime where  only few 1D continua are opened, the product ${\eta \tau}$ is small as compared to unity. In this regime, the function ${J_1(\eta,\tau)}$ can be approximated by its zero-range limit. From Eq.~\eqref{eq:J_series}, one finds
\begin{equation}
J_1(0,\tau)
= \sqrt{\pi}\left[
\frac{1}{\sqrt{-\tau}}+\zeta\left(\frac{1}{2}\right)+\sum_{p=1}^{\infty}\left(
\frac{1}{\sqrt{p-\tau}}-\frac{1}{\sqrt{p}}
\right) \right] ,
\label{eq:J1_Zero_Range}
\end{equation}
so that ${J_1(0,\tau)= \sqrt{\pi}\zeta_H\left(1/2, -\tau\right)}$ where ${\zeta_H}$ is the Hurwitz zeta function. This coincides with the result obtained by using zero-range potential models \cite{Ber03}. 

\subsection{Monomode regime}

\subsubsection{Scattering and resonance}
 
The monomode regime occurs for ${\hbar\omega_\perp \le E_{\rm rel}<3\hbar\omega_\perp}$. At large interatomic distances ${(z \gg \epsilon, a_\perp)}$, the relative particle occupies only the lowest transverse mode ${(n=0,m=0)}$ and one can define the 1D scattering amplitude ${f_{\rm 1D}}$ from the behavior of the scattering wave function
\begin{multline}
\langle z,\mathbf r_\perp|\Psi_{\rm rel}^{k_0,0,0}\rangle   \sim 
\langle \mathbf r_\perp | 0,0 \rangle
\left[ e^{i k_0 z} + f_{\rm 1D}(k_0)e^{i k_0  |z|}\right] .
\label{eq:1D_scatt_lowE} 
\end{multline}
Near the threshold of the propagating regime ${\left(E_{\rm rel} \sim \hbar\omega_\perp \right)}$, the scattering amplitude can be approximated at the effective range level:
\begin{equation}
f_{\rm 1D}(k_0) = -\frac{1}{1+ik_0 a_{\rm 1D}-i k_0^3 {R_{\rm 1D}}^3} .
\label{eq:f1D_effective_range}
\end{equation}
In Eq.~\eqref{eq:f1D_effective_range}, ${a_{\rm 1D}}$ is the 1D scattering length and ${R_{\rm 1D}}$ is the 1D-effective range parameter. They are given by:
\begin{align}
&a_{\rm 1D} =\frac{-a_\perp(a-a^{\rm TG})}{2(a-a_1^{\rm div})}
\left[\frac{a_\perp a_1^{\rm div}(1+\eta)^2}{a_{\rm bg}^2}+ \frac{\mathcal C_0^{\rm 1D}(\eta)}{\sqrt{\pi}}\right], \label{eq:a1D}\\
&{R_{\rm 1D}}^3= \frac{a_\perp^6 (1+\eta)^2}{8 R^\star a_{\rm bg}^2} \left(\frac{a/a_{\rm bg}-1}{a/a_1^{\rm div}-1}\right)^2 + \frac{C_1^{\rm 1D}(\eta)a_\perp^3}{8\sqrt{\pi}} \nonumber\\
&\qquad \qquad
+\frac{e^{-2\eta} a_\perp^2 \epsilon^2(1+\eta)^2}{4 a_{\rm eff}(\hbar \omega_\perp)} ,
\label{eq:R1D}
\end{align}
where we have introduced the two parameters ${a_1^{\rm div}}$ and ${a^{\rm TG}}$ defined by:
\begin{align}
&a_1^{\rm div} = \frac{a_{\rm bg}}{1+ {\delta \mathcal M \Delta B}/{\hbar \omega_\perp}},\\
&a^{\rm TG} = \frac
{ a_\perp 
\left( R^\star a_{\rm bg}-{a_{\rm \perp}^2}/{2} \right) + \frac{\mathcal C^{\rm 1D}_0(\eta) }
{( 1+\eta)^2 \sqrt {\pi}}R^\star a_{\rm bg}^2  
}{R^\star a_{\rm \perp}+ \frac{{\mathcal C^{\rm 1D}_0(\eta)}
}{( 1+\eta)^2 \sqrt {\pi}}\left( R^\star a_{\rm bg}+{ a_{\rm \perp}}^2/2 \right)}.
\end{align}
In the limit where ${R^\star \to 0}$ and ${\epsilon \to 0}$, one finds from Eq.~\eqref{eq:a1D} the result of Ref.~\cite{Ols98}, i.e.,
\begin{equation}
{a_{\rm 1D} \sim \frac{-a_\perp}{2}\left[\zeta\left(\frac{1}{2} \right) + \frac{a_\perp}{a} \right]}. 
\label{eq:Ols98}
\end{equation}
For a 3D scattering length such that ${a=a_1^{\rm div}}$, the 1D scattering length ${a_{\rm 1D}}$ diverges. The resonant condition in 1D, i.e. the regime where the scattering amplitude is maximum, occurs for ${a=a^{\rm TG}}$ where ${a_{\rm 1D}=0}$. In the many-boson problem, if one neglects the effective range term, the limit ${a_{\rm 1D}\to 0^-}$ gives the Tonks-Girardeau (TG) regime and the limit ${a_{\rm 1D} \to 0^+}$ gives the super Tonks-Girardeau (STG) regime \cite{Gir60,Ast05}.

\subsubsection{1D effective range model}

Similarly to the 2D-ERM, one can define a 1D effective range model (1D-ERM) which is a zero-range potential approach which gives the exact expression of the scattering amplitude in Eq.~\eqref{eq:f1D_effective_range}. In the 1D-ERM, the strictly 1D-wave function for the relative motion (denoted ${\psi_{\rm 1D}}$) depends only on the relative coordinate ${z}$ and solves the free Schr\"{o}dinger equation everywhere in the space excepted at the contact ${(z=0)}$. We introduce the projection operator over even 1D wave functions, denoted by ${\hat{\Pi}_s^{\rm 1D}}$:
\begin{equation}
\langle z | \hat{\Pi}_s^{\rm 1D} |\psi_{\rm 1D} \rangle = \frac{\psi_{\rm 1D}(z)+\psi_{\rm 1D}(-z)}{2} .
\end{equation}
In this approach, the interatomic forces are replaced by the following contact condition: 
\begin{equation}
\lim_{z \to 0^+} \left( 1 + a_{\rm 1D} \partial_z + {R_{\rm 1D}}^3 \partial_z^3\right) \langle z | \hat{\Pi}_s^{\rm 1D} |\Psi\rangle  = 0 .
\label{eq:contact_1D}
\end{equation}
The 1D-ERM generalizes the contact condition of the 1D Wigner-Bethe-Peierls (1D-WBP) model where ${R_{\rm 1D}=0}$ which defines the interaction in the Lieb-Liniger model of the 1D Bose gas \cite{Lie63}. 

\section{Bound states in atomic waveguides}

\label{sec:bound-states}

\subsection{Effect of confinement}

\begin{figure}[b]
\includegraphics[width=8cm]{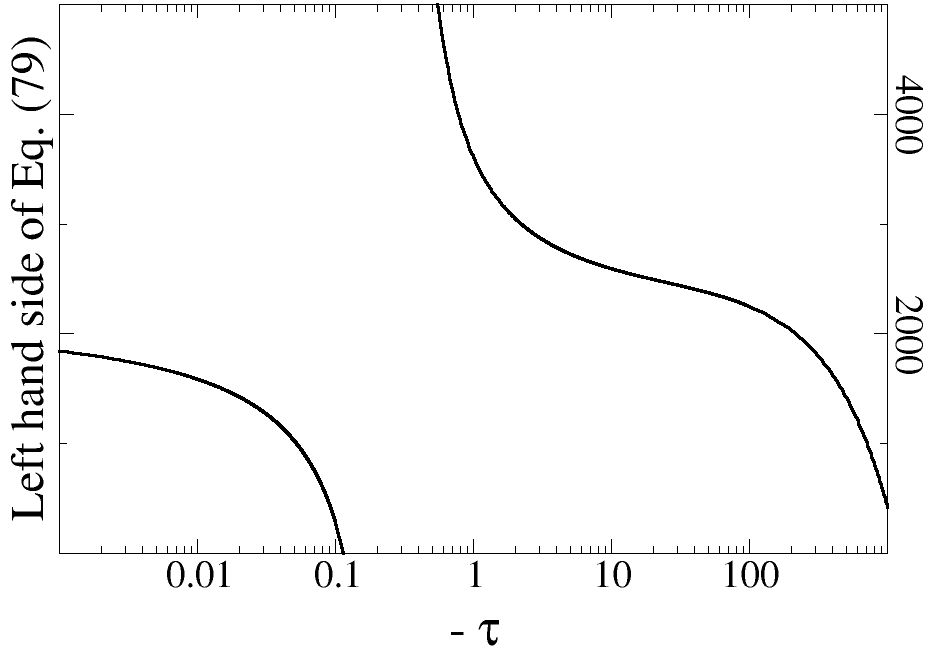}
\caption{Left-hand side of Eq.~\eqref{eq:binding_eq_D}, in a 2D atomic waveguide, using the parameters of the Cs FR from Tab.~(\ref{Tab:species}) and ${a_\perp=1000a_0}$.}
\label{fig:dimer_D}
\end{figure}

\begin{figure}[t]
\includegraphics[width=8cm]{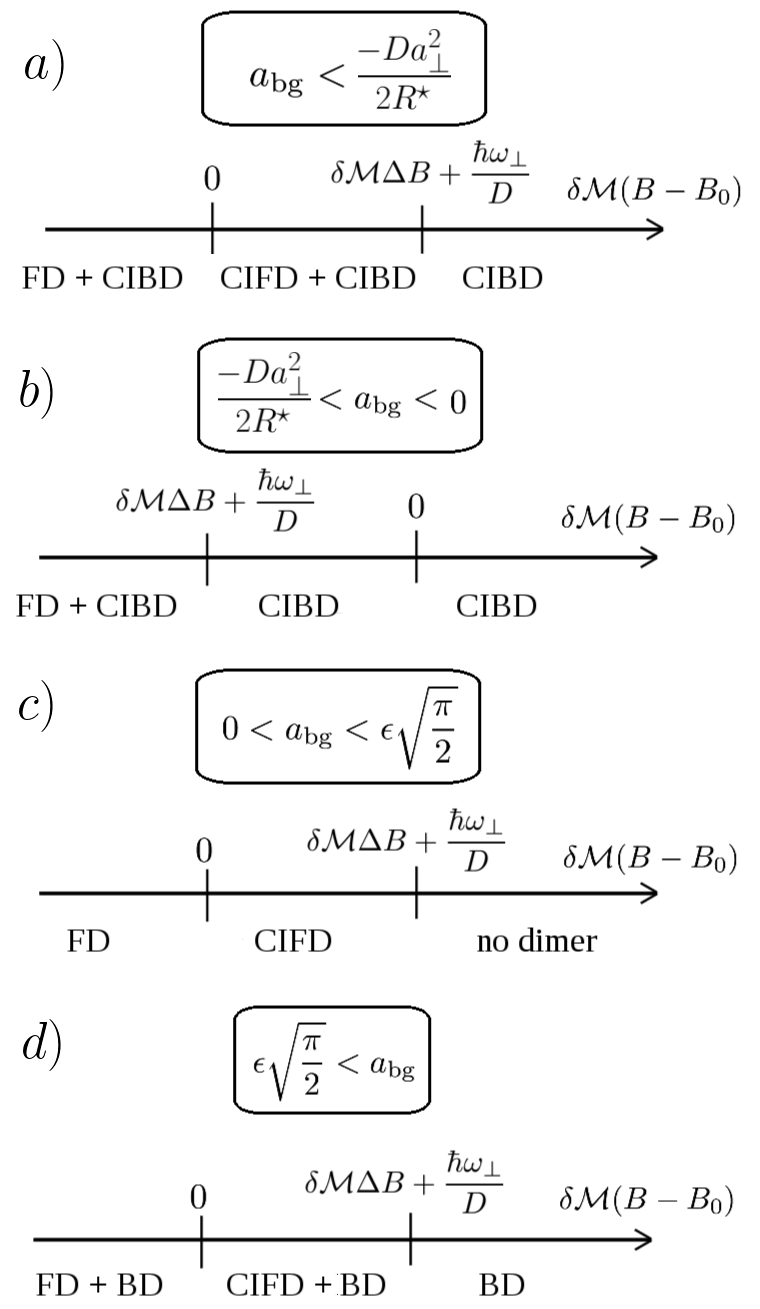}
\caption{Number and nature of dimers in reduced dimension depending on the value of ${a_{\rm bg}}$ and of the detuning ${\delta \mathcal M (B-B_0)}$. The Feshbach dimer (FD) and background dimer (BD) denote dimers that exist in the free space [see Table~\eqref{tab:dimers_3D}]. In situations where the transverse trap extends the domain of existence of the FD (or BD) dimer, we have introduced the notion of confinement-induced Feshbach dimer (CIFD) [or confinement-induced background dimer (CIBD)]. In the case ${b)}$, the domain of existence of the FD is reduced as compared to the 3D case.}
\label{fig:plot_D}
\end{figure}

We now consider the consequence of the transverse confinement on the properties of the dimers. The two main results of this part are as follows: ${(i)}$ the threshold where the FD appears is shifted at a magnetic field ${B_0^D}$: 
\begin{equation}
B_0^D = B_0 + \Delta B+ \frac{\hbar \omega_\perp}{D\delta \mathcal M} ,
\label{eq:B_0^D}
\end{equation} 
${(ii)}$ a background dimer is induced by the confinement for negative values of the background scattering length. When the domain of existence of the FD (or BD) is extended as a consequence of the confinement, we introduce the notion of confinement-induced Feshbach dimer (CIFD) [or of confinement-induced background dimer (CIBD)]. 

In presence of a transverse harmonic trap, the energy of a bound state is below the threshold of the first continuum, i.e., ${E<\frac{\hbar\omega_\perp}{D}}$, where  ${D=2}$ in a 2D atomic waveguide and ${D=1}$ for a 1D atomic waveguide. In what follows, we use the reduced energy ${\tau}$ defined in Eq.~\eqref{eq:def_tau} with ${\lambda=E}$. For the dimers, the reduced energy is thus negative. We also define the binding energy of a bound state ${E_{\rm dim}}$ in a $D$-dimensional waveguide with respect to the threshold of the monomode regime:
\begin{equation}
E_{\rm dim} = E - \frac{\hbar \omega_\perp}{D} .
\end{equation}
The spectrum of the dimers is given by the poles of the transition matrix in Eq.~\eqref{eq:transition_matrix_2D} or~\eqref{eq:transition_matrix_1D} which verify:
\begin{align}
&\left(2\tau+\frac{1}{D}\right) + \frac{\delta \mathcal M  \Delta B}{\hbar \omega_\perp} 
\frac{\mathcal F_D(\eta,\tau)}
{\mathcal F_D(\eta,\tau)-1} =  \frac{\delta \mathcal M  (B-B_0)}{\hbar \omega_\perp}
\label{eq:binding_eq_D},\\
&\text{where} \quad
\mathcal F_{D}(\eta,\tau) = \frac{-a_{\rm bg}e^{-4\eta\tau} J_D(\eta,\tau)}{a_\perp\sqrt{\pi} (1+\eta)^{3-D} }.
\end{align}
The left-hand of Eq.~\eqref{eq:binding_eq_D} is an increasing function of ${\tau}$ which varies from ${-\infty}$ for ${\tau \to -\infty}$ to ${\left(\frac{1}{D} + \frac{\delta \mathcal M  \Delta B}{\hbar \omega_\perp }\right)}$ for  ${\tau \to 0^-}$. This last limit gives the threshold in Eq.~\eqref{eq:B_0^D} at which the CIFD appears. For ${0<a_{\rm bg}<\epsilon \sqrt{\frac{\pi}{2}}}$, it is a continuous function in the interval ${\tau \in ]-\infty,0[}$ and Eq.~\eqref{eq:binding_eq_D} admits thus at most one solution. In the other cases, the left-hand side of Eq.~\eqref{eq:binding_eq_D} diverges for a given negative value of the reduced energy ${\tau}$. As a consequence, depending on the value of ${a_{\rm bg}}$ and of the detuning ${\delta \mathcal M (B-B_0)}$, there are one or two dimers. The behavior of the left-hand side of Eq.~\eqref{eq:binding_eq_D} in this last regime is illustrated in Fig.~\eqref{fig:dimer_D} using the example of the Cs FR in Table~\eqref{Tab:species}. We have summarized the domain of existence of the dimers in Fig.~\eqref{fig:plot_D}. For standard FR studied experimentally, the ratio ${|{\delta \mathcal M \Delta B}|/{\hbar \omega_\perp}={a_\perp^2}/{(2 R^\star |a_{\rm bg}|)}}$ is very large and ${|{\delta \mathcal M  \Delta B}|>{\hbar \omega_\perp}/D}$. Consequently, ${\delta \mathcal M  \Delta B +\hbar \omega_\perp/D }$ has the same sign as ${a_{\rm bg}}$ [see Eq.~\eqref{eq:DeltaB}]. However, for ultra-narrow resonances where the ratio ${|{\delta \mathcal M  \Delta B}|/{\hbar \omega_\perp}}$ is very small, ${\delta \mathcal M  \Delta B +\hbar \omega_\perp/D }$ is thus positive. In this particular case, the domain of existence of the dimers is thus modified with respect to standard FR for ${a_{\rm bg}<0}$: there is still a CIBD whatever the value of the energy detuning, but also a CIFD exists in the interval 
${0<\delta \mathcal M (B-B_0)<\delta \mathcal M \Delta B+ {\hbar \omega_\perp}/{D}}$ [case ${a)}$ of Fig.~(\ref{fig:plot_D})].

In the limit of large reduced binding energies, one recovers the expected 3D result of Eq.~\eqref{eq:Edim_large_detuning}.

\subsection{Shallow bound states and zero-range approach}

The bound states can be considered shallow in the atomic waveguides for small reduced binding energies ${|\tau | \ll 1}$. Consequently, they can be studied in the monomode regime by using zero-range models.

\subsubsection{2D geometry}

In the 2D geometry, the scattering length ${a_{\rm 2D}}$ is always positive and thus the 2D-WBP model admits always one single bound state of binding energy
${E_{\rm dim} = E^{\rm WBP}_{D=2}}$ with
\begin{equation}
E^{\rm WBP}_{2} =-\frac{2\hbar^2 e^{-2\gamma}}{\mu a_{\rm 2D}^2} .
\label{eq:dim_2D_WBP}
\end{equation}
Equation \eqref{eq:dim_2D_WBP} can be generalized by considering an energy-dependent 2D scattering length ${a_2(E_{\rm rel})}$ \cite{Kan06}. This last approach considered with
${a_2(E_{\rm rel})=a_{\rm 2D} \exp(-2\mu a_{\rm 2D} R_{\rm 2D}E_{\rm rel}/\hbar^2)}$, is equivalent to the 2D-ERM and the binding energy ${E_{\rm dim}}$ of the 2D bound state is then found by solving
\begin{equation}
\ln \left(\frac{E_{\rm dim} }{E^{\rm WBP}_{2}}\right)= \frac{-8 R_{\rm 2D}e^{-2\gamma}}{a_{\rm 2D}}  \frac{E_{\rm dim} }{E^{\rm WBP}_{2}}.
\label{eq:dim_2D_ERM}
\end{equation}
In Eqs.~\eqref{eq:dim_2D_WBP} and~\eqref{eq:dim_2D_ERM}, ${|E_{\rm dim}| \ll \hbar \omega_\perp}$ for large values of the 2D scattering length (${a_{\rm 2D}\gg a_\perp}$ ), i.e., in the resonant regime, as is the case for example near the threshold of appearance of the CIFD.

\begin{figure}[t]
\includegraphics[width=7.8cm]{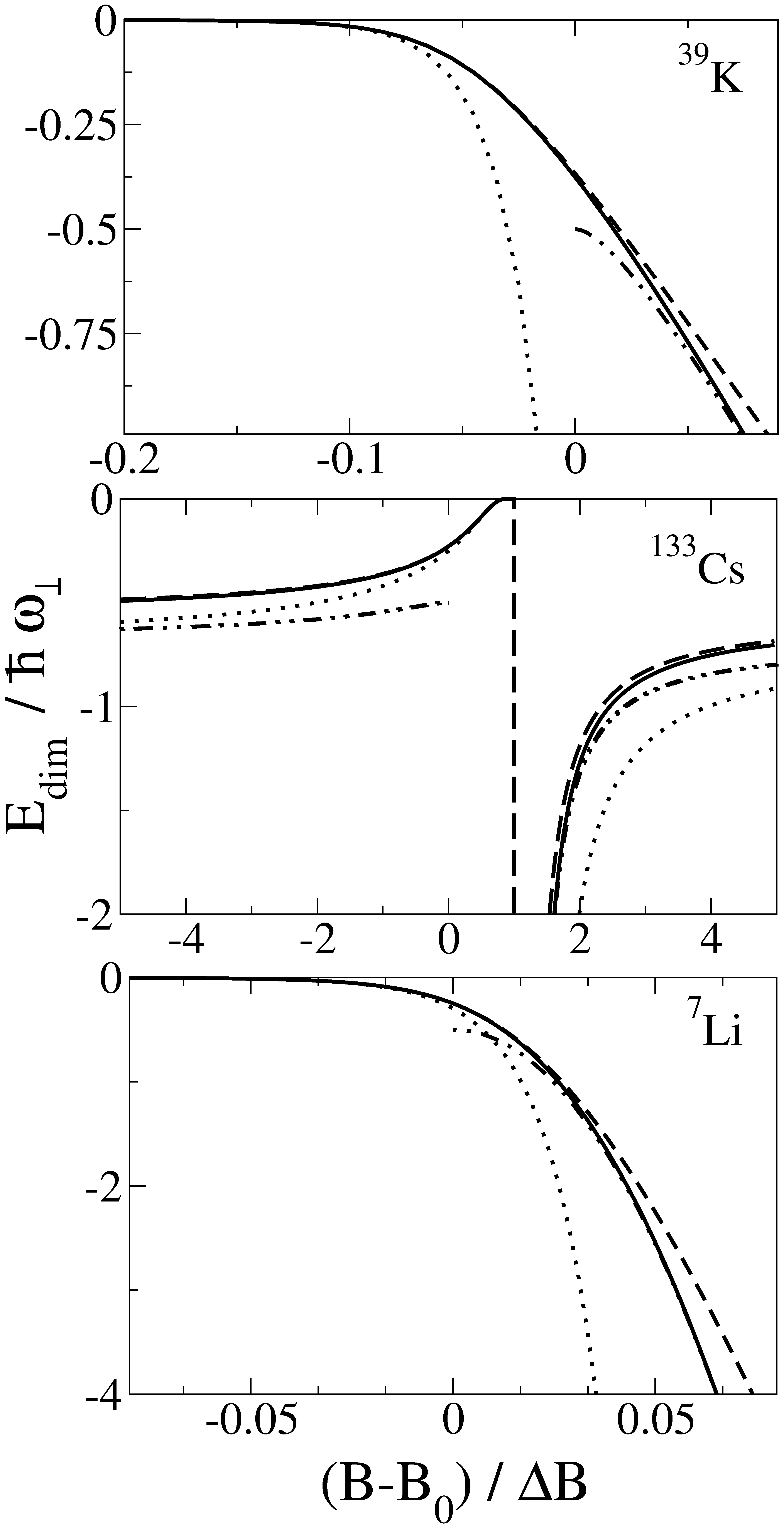}
\caption{Spectrum of quasi-2D dimers for three FR of Table~\eqref{Tab:species} with ${a_\perp=1000a_0}$. Solid line: two-channel model; dotted line: 2D-WBP model; dashed line: 2D-ERM; double dotted dashed line: 3D spectrum. For $^{39}$K and $^{7}$Li,  only one dimer branch can be seen. The other branch is very close to zero and exists for ${(B-B_0)/\Delta B \gtrsim 1}$. When ${\delta \mathcal M (B-B_0)}$ tends to ${\delta \mathcal M \Delta B+ \frac{\hbar \omega_\perp}{D}}$ by positive values,  the two-dimensional scattering length tends to zero and there is no shallow dimer in the two-channel model. Nevertheless, in this regime the quasi-vertical dashed line in the Cs case illustrates the fact that the 2D-ERM predicts a spurious dimer in a tiny interval of the relative detuning.}
\label{fig:dim_2D}
\end{figure}
Examples of binding energies for different FR of Table~\eqref{Tab:species} are plotted in Fig.~\eqref{fig:dim_2D} using the two-channel model, the WBP model, and the 2D-ERM. We have chosen a common transverse length  ${(a_\perp=1000 a_0)}$ for all species. The Cs resonance has been intensively studied in the group of Insbr\"{u}ck and has been used for the first achievement of 2D BEC with ultracold atoms \cite{Ryc04}. It illustrates a FR in the vicinity of a shape resonance (${|a_{\rm bg}| \gg \epsilon}$) with a shallow background dimer in the 3D free space. The FR chosen for $^{39}$K illustrates a narrow FR (${R^\star \gg |a_{\rm bg}|}$) and for $^{7}$Li it illustrates a standard broad FR. In these last two examples, the background scattering length is negative, and there is thus a CIBD. Moreover, the scattering length is small compared to the transverse length, and in a large interval of detuning the ratio ${a_\perp / a_{\rm eff}(\hbar \omega_\perp/2) \sim a_\perp/a_{\rm bg}}$ is large and negative and the 2D scattering length in Eq.~\eqref{eq:a2D} is thus extremely large due to the exponential law. Consequently in a large interval of detuning, the CIBD is an ultra shallow bound state with ${|E_{\rm dim}| \ll \hbar \omega_\perp}$. The three resonances admit one or two bound states depending on the detuning. 
\subsubsection{1D geometry}

The 1D-WBP model supports one bound state iff the 1D scattering length is positive ${a_{\rm 1D}>0}$ with the binding energy ${E_{\rm dim} = E^{\rm WBP}_{D=1}}$: 
\begin{equation}
E^{\rm WBP}_{1} =-\frac{\hbar^2}{2\mu a_{\rm 1D}^2} .
\label{eq:Edim_WBP_1D}
\end{equation}
\begin{figure}[b]
\includegraphics[width=8cm]{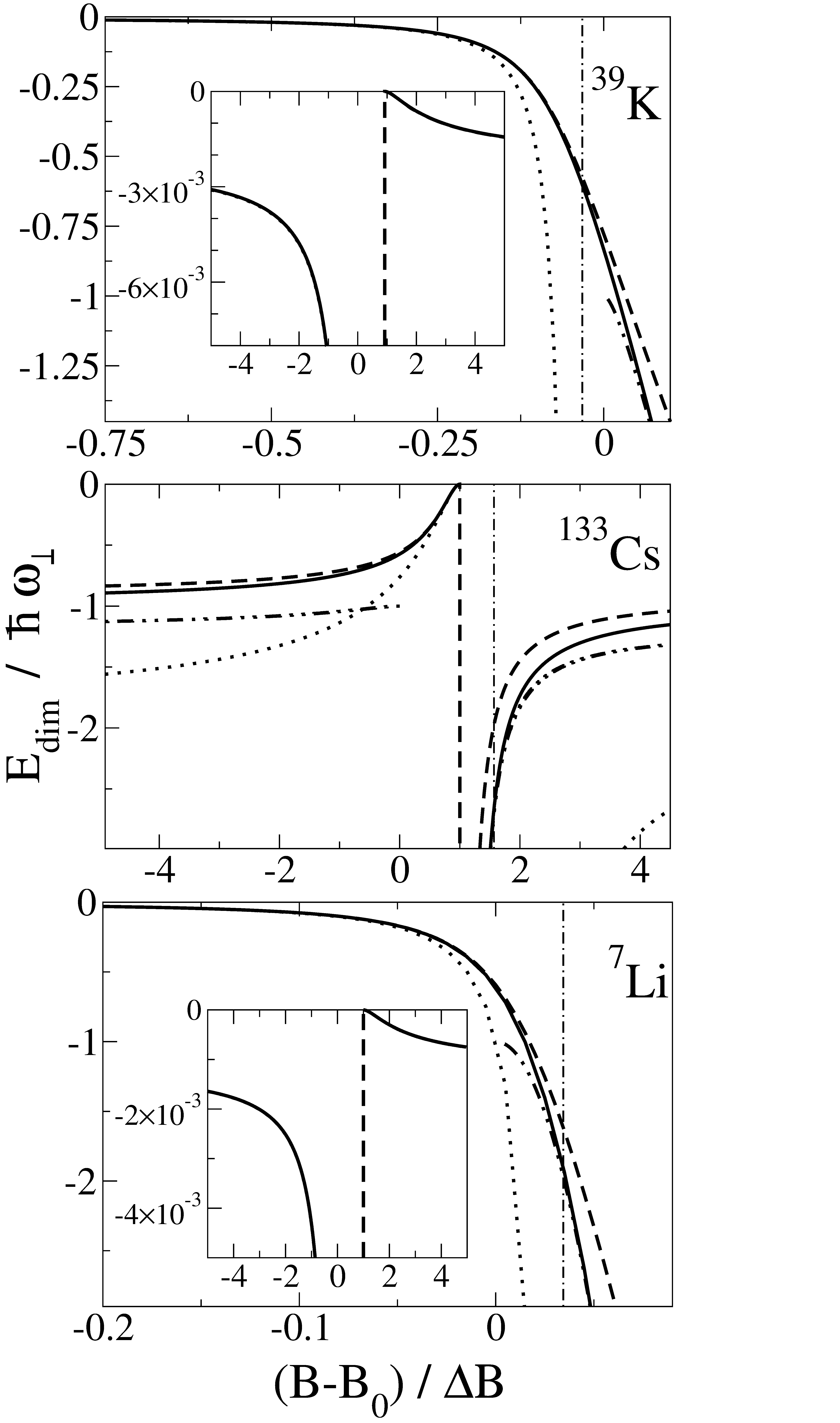}
\caption{Spectrum of quasi-1D dimers for three FR of Table~\eqref{Tab:species} with ${a_\perp=1000a_0}$. Solid line: two-channel model; dotted line: 1D-WBP model; dashed line: 1D-ERM; double dotted dashed line: 3D spectrum. In the inserts, the low-dimensional predictions are too close to be distinguished. The vertical dashed-dotted lines indicate the detuning where the 1D resonance ${(a_{\rm 1D}=0)}$ occurs. As in the 2D case, in a tiny interval of the relative detuning where ${\delta \mathcal M (B-B_0)}$ tends to ${\delta \mathcal M \Delta B+ \frac{\hbar \omega_\perp}{D}}$ by positive values,  the quasi-vertical dashed lines illustrate the fact the 1D-ERM  predicts the existence of a spurious dimer branch.}
\label{fig:dim_1D}
\end{figure}
Referring to the usual pairwise pseudo-potential ${g_{\rm eff} \delta(z)}$ used in 1D systems, one has ${g_{\rm eff}= - \hbar^2/(\mu a_{\rm 1D})}$ and Eq.~\eqref{eq:Edim_WBP_1D} is thus relevant in the regime of small and attractive effective interaction (${g_{\rm eff} \to 0^-}$).  The 1D-ERM allows zero, one, or two bound states, depending on the values of the parameters ${R_{\rm 1D}}$ and $a_{\rm 1D}$. The binding energies $E_{\rm dim}={-\hbar^2q^2}/{(2\mu)}$ are given by the positive roots of the equation
\begin{equation}
1 - a_{\rm 1D} q - R_{\rm 1D}^3 q^3 = 0.
\label{eq:polynome_1D}
\end{equation}
We plotted in Fig.~\eqref{fig:dim_1D} the binding energies in the 1D atomic waveguide for three different resonances of Tab.~\eqref{Tab:species}  and  ${a_\perp=1000 a_0}$. This figure permits to compare  the binding energy of shallow dimers obtained from the two-channel model,  the 1D-WBP model and the 1D-ERM.  In the 1D resonant regime where the 1D scattering length tends to zero the WBP model cannot describe a possible shallow dimer whereas the 1D-ERM can give a quantitative approximation [see Fig.~\eqref{fig:dim_1D}]. 

\subsection{Relevance of the effective range approximation in low dimensions}

We first investigate the regime where the effective range term gives a perturbative contribution to the bound state energy in the monomode regime for ${D=1}$~or~${2}$: ${E_{\rm dim} = E^{\rm WBP}_D + \delta E_D}$. Linearizing Eqs.~\eqref{eq:polynome_1D} and~\eqref{eq:dim_2D_ERM}, in the regime of large and positive value of ${a_{\rm 1D}/R_{\rm 1D}}$ or ${a_{\rm 2D}/R_{\rm 2D}}$, one finds:
\begin{align}
&\frac{\delta E_{D=1}}{E^{\rm WBP}_{D=1}} = -\frac{2}{3+\left(\frac{a_{\rm 1D}}{R_{\rm 1D}}\right)^3}, 
\label{eq:approx_bound_1D}\\
&\frac{\delta E_{D=2}}{E^{\rm WBP}_{D=2}} = -\frac{1}{1+ \frac{e^{2\gamma}}{8} \left(\frac{a_{\rm 2D}}{R_{\rm 2D}}\right) }.
\label{eq:approx_bound_2D}
\end{align}
When the detuning ${\delta \mathcal M (B-B_0)}$ tends to ${\delta \mathcal M \Delta B+ \frac{\hbar \omega_\perp}{D}}$ by positive values, for ${D=2}$, ${a_{\rm 2D}}$ tends to zero, whereas for ${D=1}$, ${a_{\rm 1D}}$ tends to minus infinity. In this regime, no shallow dimer exists in the two-channel model and in the WBP model. Nevertheless, the ERM predicts a spurious binding energy in a narrow interval of the relative magnetic detuning. This can be seen in Figs.~(\ref{fig:dim_2D}) and~(\ref{fig:dim_1D}) by the existence of a quasi-vertical dashed line in the spectrum. 
Interestingly, as a consequence of the non trivial dependence of the effective range parameters ${R_{\rm 1D}}$ and ${R_{\rm 2D}}$ as a function of ${R^\star}$, the 1D-ERM and 2D-ERM are much more predictive than the WBP models for binding energies of the shallow 1D bound states even for broad FR where the 3D effective range parameter is small [for instance, for the Cs FR in Figs.~\eqref{fig:dim_2D} and~\eqref{fig:dim_1D}]. Moreover, the 1D-ERM and 2D-ERM provide still quantitative results for binding energies of the order of the trapping energy ${\hbar\omega_\perp}$. 

We now identify in the 1D atomic waveguide a new regime where the effective range approximation is relevant non perturbatively. This happens in the resonant limit, i.e. for ${a \sim a^{\rm TG}}$ where ${a_{\rm 1D} \to 0}$ and for a large effective range parameter ${R_{\rm 1D}}$:
\begin{equation}
R_{\rm 1D} \gg a_\perp  \quad \text{and} \quad |a_{\rm 1D}| \ll a_\perp .
\label{eq:condition_0}
\end{equation}
When Eq.~\eqref{eq:condition_0} is satisfied, the effective range term can be anomalously large with ${k |a_{\rm 1D}| \ll  (k|R_{\rm 1D}|)^3}$ in the monomode regime and for small relative momentum ${(k \ll 1/a_\perp)}$ (typically for ${k \sim 1/R_{\rm 1D}}$). In the strict resonant limit (${a_{\rm 1D} =0}$) and for positive value of ${R_{\rm 1D}}$, there is a shallow dimer of binding energy ${\hbar^2/(2\mu R_{\rm 1D}^2)}$. This is in strong contrast with the result  of the WBP model where the dimer is infinitely bound in the STG limit ${(a_{\rm 1D} \to 0^+)}$ and does not exist in the TG limit ${(a_{\rm 1D} \to 0^-)}$. Equation \eqref{eq:condition_0} defines a regime which has not yet been explored at the many-body level. In this context, it is interesting to address more closely the validity of the effective range approximation for the binding energy  ${|k| \sim 1/R_{\rm 1D}}$. For this purpose, one evaluates the ratio between the effective range term in the denominator in  Eq.~\eqref{eq:1D_scatt_lowE} and the next term in the low-energy expansion obtained from Eq.~\eqref{eq:transition_matrix_1D} at ${k=i/R_{\rm 1D}}$ and ${a=a^{\rm TG}}$. For ${\eta \ll 1}$, one finds the following criterion:
\begin{equation}
\left|\frac{a_{\rm bg}}{a_\perp}\left( \frac{R^\star}{a_\perp + a_{\rm bg} \zeta(1/2)  }\right) ^{1/3} \right|  \ll 1.
\label{eq:condition_1}
\end{equation}
Moreover, in view of a future achievement of this regime, it is essential to know the necessary conditions to reach large values of the ratio ${R^\star/a_\perp}$. In the TG regime [where the second condition in Eq.~\eqref{eq:condition_0} is satisfied] and neglecting $\eta$, Eq.~\eqref{eq:R1D} gives
\begin{equation}
\left(\frac{R_{\rm 1D}}{a_\perp}\right)_{a=a^{\rm TG}} ^{3} \simeq \frac{R^\star}{2a_\perp} \left[ 1+\zeta(1/2) \frac{a_{\rm bg}}{a_\perp}\right]^2 .
\label{eq:largeRstar}
\end{equation}
Hence, the first condition in Eq.~\eqref{eq:condition_0} can  be satisfied for large enough values of the ratio ${R^\star/a_\perp}$. However, this ratio cannot be too large in order to satisfy Eq.~\eqref{eq:condition_1}. 
\begin{figure}[t]
\includegraphics[width=7.8cm]{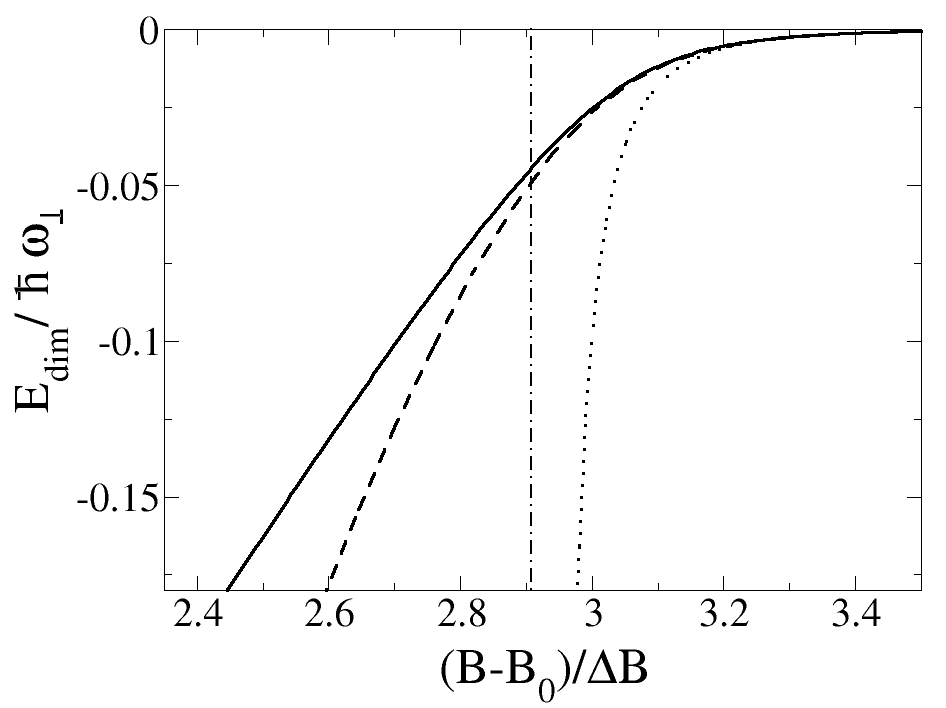}
\caption{Energies of the dimer as a function of the reduced detuning in the 1D monomode regime ($a_\perp=3000 a_0$) for the FR of $^{23}$Na  at ${B_0=853}$~G  [see Table~\eqref{Tab:species}]. Solid line: two-channel model; dotted line: 1D-WBP model; dashed line: 1D-ERM. The vertical dashed-dotted line indicates the detuning where the 1D resonance ${(a_{\rm 1D}=0)}$ occurs.}
\label{fig:Na_1D_aperp_3000}
\end{figure}
Hopefully, species which verify the two conditions in Eqs.~\eqref{eq:condition_0} and~\eqref{eq:condition_1} can be found. One example is given by $^{23}$Na near the narrow FR at ${853}$~G with a realistic axial confinement with ${a_\perp=3000 a_0}$  [see Tab.~\eqref{Tab:species}]. At resonance ${a_{\rm 1D}=0}$, one finds ${R_{\rm 1D}/a_\perp\sim 3.2}$ and the value of the left-hand side of the criterion in Eq.~\eqref{eq:condition_1} is about ${(0.087)}$. The binding energy is plotted  as a function of relative detuning in Fig.~\eqref{fig:Na_1D_aperp_3000}. Close to the resonance, the dimer is shallow and this plot illustrates the breakdown of the WBP model in this regime. As expected, the ERM model gives a quantitatively good approximation of the binding energy.

\subsection{Occupation probability of the closed channel}
\label{sec:occupation_proba}

The mean number of molecules in the closed channel highlights the crossover between the different types of dimers. It is equal here to the occupation probability of the molecular state ${\langle \Psi^{\rm m} |\Psi^{\rm m} \rangle= \frac{dE}{dB}/{\delta \mathcal M} }$ \cite{Wer09}. In the off-resonant regime, for the BD the occupation probability in the closed channel vanishes, whereas for the FD its tends to unity (for large detunings and thus large binding energies, the FD populates mainly the molecular state).

\begin{figure}[hx]
\includegraphics[width=8cm]{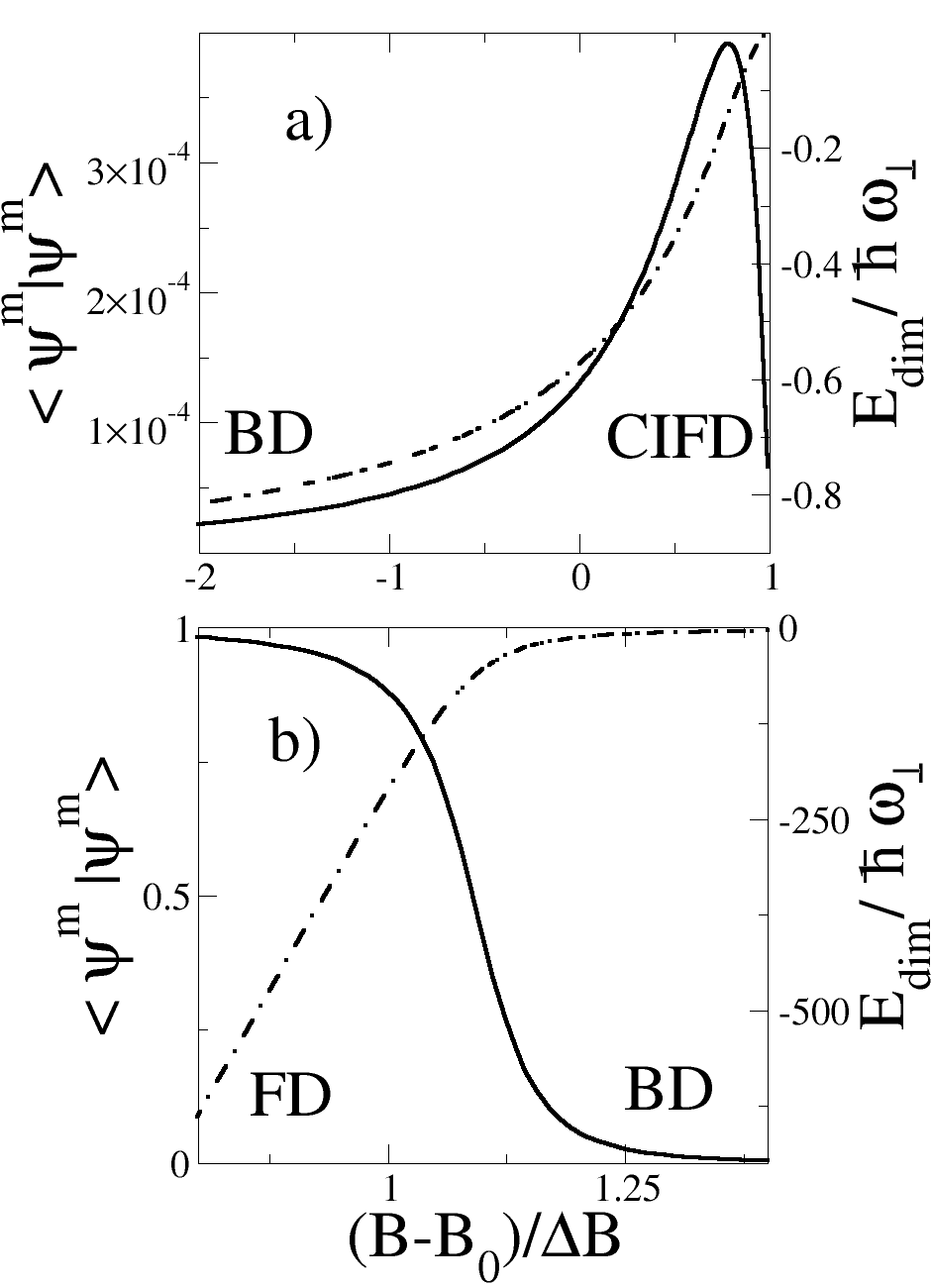}
\caption{Dashed-dotted lines: binding energy for dimers in a 1D atomic waveguide (${a_\perp=1000 a_0}$) for the $^{133}$Cs FR in Table~\eqref{Tab:species}; solid lines: occupation probability of the closed channel for the corresponding dimer. ${a)}$ Results for the dimers of the upper branch in the spectrum; ${b)}$ results for the lower branch.}
\label{fig:proba_cc}
\end{figure}

We plotted the occupation probability in the closed channel  in Fig.~\eqref{fig:proba_cc} for the two bound states in a 1D atomic waveguide for the $^{133}$Cs FR at ${B_0=-11.7}$~G with a transverse length ${a_\perp=1000 a_0}$.

In this example, the background scattering length ${a_{\rm bg}}$ is positive and larger than ${\epsilon\sqrt{\frac{\pi}{2}}}$: this resonance corresponds to the case d) in Fig.~(\ref{fig:plot_D}). The figure illustrates the crossover between the CIFD and the BD [Fig.~(\ref{fig:proba_cc}-${a)}$] and the BD and the FD [Fig.~(\ref{fig:proba_cc}-${b)}$]. At the threshold where the CIFD appears, the occupation probability in the closed channel is small. In the off-resonant regime, the BD has also a vanishingly small occupation probability in the closed channel. Conversely, for large and negative magnetic detuning, the FD occupies mainly the closed channel as expected from~Eq.~\eqref{eq:Edim_large_detuning}.

\section*{CONCLUSIONS}

In this paper, we use a finite-range two-channel model to solve the two-body problem for atoms in a 1D and a 2D atomic waveguide. The model which was already used in Refs.~\cite{Jon10,Mor11a,Pri11c,Tre12,Wer09}, gives quantitative results in a large interval of magnetic detuning and colliding energies and permits us to derive analytical expressions for two-body scattering in low dimensions. The model gives substantial improvement with respect to the pure WBP models of the interaction \cite{Pri11a}. We compare the results of the two-channel model with purely 2D (or 1D) zero-range models (WBP models and effective range models), and give conditions where these approaches are relevant. The effective range model is shown to be a quantitative approach in a large interval of magnetic detuning. For 1D atomic waveguides, we identify a regime of scattering resonance (the 1D scattering length tends to zero) where the effective range term is dominant in the 1D scattering amplitude. We exhibit a possible achievement of this regime by using a narrow FR of  $^{23}$Na atoms. In this regime, contrary to the standard 1D-resonant regime, a shallow dimer exists. The consequences of these findings at the few- and many-body levels are an open issue. One expects large deviations from the TG and STG properties as for instance the nature ground state \cite{McG64,Cas09} as a function of the effective range parameter.

\section*{ACKNOWLEDGEMENT}

The Cold Atoms group of LPTMC is associated with Institut Francilien de Recherche sur les Atomes Froids.

\appendix

\section*{Appendix: A USEFUL EXPRESSION}

\label{ap:Thesum}

In this appendix, we give an analytical expression for the infinite sum
\begin{equation}
S_\eta(u) = \sum_{n=0}^{\infty} | \langle \delta_\epsilon^{\rm 1D} | n \rangle|^2 \times e^{-n u/2} .
\label{eq:Seta_def}
\end{equation}
Formally, one recognizes in Eq.~\eqref{eq:Seta_def} the density operator ${\hat{\rho}}$, of the harmonic 
oscillator in the $x$-direction, considered in the representation ${|n\rangle}$ at a temperature 
${k_B T=2\hbar \omega_\perp/u}$:
\begin{equation}
S_\eta(u)=  \langle \delta_\epsilon^{\rm 1D} |\hat{\rho} | \delta_\epsilon^{\rm 1D} \rangle \times e^{u/4} .
\label{eq:mapping}
\end{equation}
We perform the calculation in the configuration space, where the density matrix ${\rho(x,x')\equiv\langle x | \hat{\rho}|x'\rangle}$ is
\begin{equation}
\rho(x,x') = \langle x | \exp\left[-\left(\frac{\hat{p}^2}{2\mu} + \frac{\mu}{2} \omega_\perp^2 \hat{x}^2 \right)\times \frac{u}{2\hbar \omega_\perp} \right] |x' \rangle .
\end{equation}
One can then use the known result for the density matrix:
\begin{equation}
\rho(x,x') = \frac{ \exp\left\{-\frac{\left[ \cosh\left(\frac{u}{2}\right)(x^2+x'\,^2)-
2 x x' \right]}{2 a_\perp^2 \sinh\left(\frac{u}{2}\right)} \right\}  }{a_\perp \sqrt{2\pi \sinh\left(\frac{u}{2}\right)}}
\end{equation}
together with the expression of  ${\langle x| \delta_\epsilon^{\rm 1D}\rangle}$ in Eq.~\eqref{eq:delta_epsilon_x}. The integration over $x$ and $x'$ implies only Gaussian functions and one finds
\begin{equation}
S_\eta(u) =\frac{1}{a_\perp \sqrt{\pi}} \frac{1}{{\sqrt{(1+ \eta)^2 - e^{-u} (1-\eta)^2}}}
\label{eq:Seta}
\end{equation}


\begin{thebibliography}{99}
\bibitem{convert} The coupling factor $\Lambda$ in the present paper differs by a factor $\sqrt{2}$ from the one in Refs.~\cite{Jon10,Pri11c}. The other parameters are ${g=g_0}$, ${{\delta \mathcal M}=\delta \mu}$ and ${\epsilon=b \sqrt{2}}$.
\end{thebibliography}
\end{document}